\documentclass[12pt,preprint]{aastex}

\shorttitle{Open Clusters White Dwarfs}
\shortauthors{Jeffery, et al.}

\begin{document}

\title{New Techniques to Determine Ages of Open Clusters Using White Dwarfs\footnotemark\footnotetext{WIYN Open Cluster Study XXX}}

\author{E. J. Jeffery, T. von Hippel, W. H. Jefferys, D. E. Winget, N. Stein, and S. DeGennaro}
\affil{Astronomy Department, University of Texas at Austin,
    Austin, TX 78712}
\email{ejeffery@astro.as.utexas.edu}

\begin{abstract}
   Currently there are two main techniques for independently determining the
ages of stellar populations: main sequence evolution theory (via cluster
isochrones) and white dwarf cooling theory.  Open clusters provide the
ideal environment for the calibration of these two clocks.  Because current 
techniques to derive cluster ages from white dwarfs are observationally 
challenging, we discuss the feasibility of determining white dwarf ages from 
the brighter white dwarfs alone.  This would eliminate the requirement of 
observing the coolest (i.e., faintest) white dwarfs.  We discuss our method 
for testing this new idea, as well as the required photometric precision and 
prior constraints on metallicity, distance, and reddening.  We employ a new 
Bayesian statistical technique to obtain and interpret results.
\end{abstract}


\section{Introduction}
 \label{intro}

   Age measurements are a fundamental problem in astronomy.  Understanding
the formation sequence of the Galaxy is largely dependent upon accurately
knowing the ages of its constituents.  There are two main techniques for 
independently determining the ages of stellar populations: main sequence (MS) 
evolution theory (via cluster isochrones) and white dwarf (WD) cooling 
theory.  Ages determined from the MS turnoff (MSTO) of globular clusters
provide the most reliable age of the Galactic halo (e.g., Chaboyer et al. 
1996), while WD cooling ages provide the most reliable age of the Galactic 
disk \citep{winget87,oswalt96,leggett98,knox99}.  Before ages determined by 
these two techniques can be meaningfully compared, they must be calibrated to 
the same absolute scale.  The best way to do this is to measure and compare WD 
ages and MSTO ages in several open clusters with a wide range of ages and 
metallicities.

   Cluster WD ages are determined by observing the coolest WDs in the 
cluster.  There is a simple relationship between the cooling time (which is 
comparable to its total age for the oldest WDs, given their short MS 
lifetimes) and the luminosity of a WD.  Therefore, once the coolest WDs are 
found, the age of the cluster can be determined.  The first studies to apply 
this technique in open clusters were done by \citet{clav95} and 
\citet{vgj95}.  Later studies \citep{richer98,vg00,vh01,clav02} showed good 
agreement in WD ages and MS ages for clusters up to 4 Gyr.  A summary of these 
studies and techniques has recently been presented by \citet{ted05}.

   This current technique to derive open cluster ages using WDs is  
observationally challenging.  The coolest WDs are intrinsically faint ($M_{V} 
\ge$ 16), limiting observations to clusters within a few kiloparsecs, and 
space-based observations are required for the oldest clusters, particularly to 
obtain morphological information to remove contaminating background galaxies.  
To obtain WD ages of more distant clusters (thus increasing the available 
sample for study), and to reduce the need for space-based observations, we 
have been motivated to explore new ways to obtain age information from the 
cluster WDs.  One solution is to explore what age information, if any, is 
available in the brighter cluster WDs.

   If successful, the payoff of a technique to derive ages based on only the 
bright WDs is great.  Cluster WD ages will be obtained more routinely because 
the brighter WDs are easier to observe and deep ground-based data will often 
suffice.  By relying less on space-based data, we will be able to obtain ages 
for more clusters (as obtaining spaced-based data is very competitive) which 
will push our calibration of WD and MSTO ages further.  Also, using the bright 
WDs means that we can observe clusters at greater distances, increasing the 
number of clusters available for study.  This technique would allow WDs 
to be an increasingly powerful and effective tool for determining ages.

   For reasons to be discussed in more detail in the following sections, ages 
we have derived using the bright WDs are precise, relative ages, rather than 
accurate, absolute ages.  Like other age indicators, we will need to perform 
extensive calibrations before this age indicator can be used as an absolute 
chronometer.

   This paper is organized as follows: we discuss the rationale behind and 
techniques used to explore the bright WD idea in Section \ref{tests}.  In 
Section \ref{results} we present results obtained from our modeling and 
Bayesian analysis, as well as present preliminary constraints on the precision 
and completeness required to obtain good ages.  We end with concluding remarks 
in Section \ref{conclusion}.

\section{Ages from the Bright Cluster WDs}
  \label{tests}

\begin{figure}
   \epsscale{0.95}
	\plotone{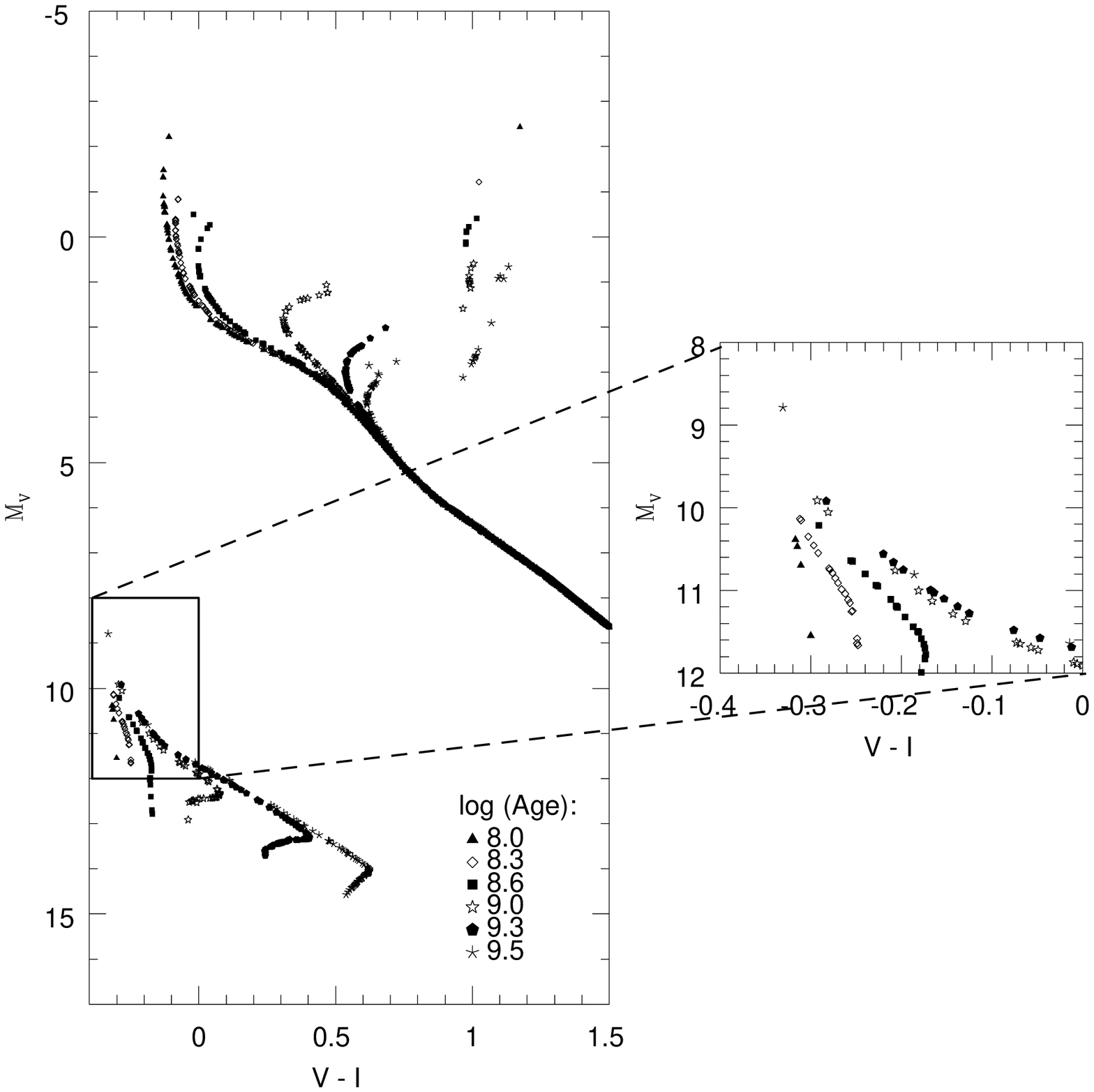}
	\caption{Simulated  clusters for several different ages.  The expanded  
		region shows the regime of the brighter WDs, clearly showing 
		the subtle differences in the slopes and positions of the WD 
		cooling	sequences relative to the MS for clusters of different 
		ages.  This makes it possible to extract age information 
		without observing the faintest WDs.}
	\label{simclus}
\end{figure}

   In Figure \ref{simclus} we present several color-magnitude diagrams (CMDs) 
for simulated clusters of varying ages, along with an expanded view of the 
region of the bright WDs.  It is clear from this figure that in the regime of 
the brighter WDs there are subtle differences in the slope and position 
relative to the MS of the WD cooling sequences for clusters of different 
ages.  These differences are what make it possible to extract age information 
without observing to the WD terminus.

\subsection{Rationale of the Bright WD Idea}
  \label{rationale}

   What is the physical reason for the slope differences of the WD cooling 
sequence in the CMD?  Assuming the mapping between a WD's mass and its MS 
counterpart's mass is universal and single-valued, younger clusters have 
higher mass TO stars, and therefore higher mass bright WDs.  The overall mass 
of a WD affects its position on the CMD; e.g., higher mass WDs have smaller 
radii and are therefore fainter.

   This mapping between a WD's mass and its MS mass is known as the 
Initial-Final Mass Relation (IFMR).  Ages determined using the traditional 
cool WD method, as described briefly in Section \ref{intro}, have little 
dependence on the IFMR.  However, the IFMR greatly affects the hot, bright 
WDs.  The shape and position of the WD cooling sequence relative to the MS in 
the bright WD regime is a mass effect.  Therefore, if there is significant 
cluster-to-cluster variation in the IFMR, the bright WD technique breaks 
down.  However, it is the general consensus among researchers that the IFMR is 
single-valued and the same cluster-to-cluster \citep{weidemann00}.  Because of 
this dependence on the IFMR, this technique is a relative age indicator.  
Once the IFMR becomes convincingly single-valued and accurately calibrated, 
this technique could yield absolute ages.

\begin{figure}
   \epsscale{0.95}
	\plotone{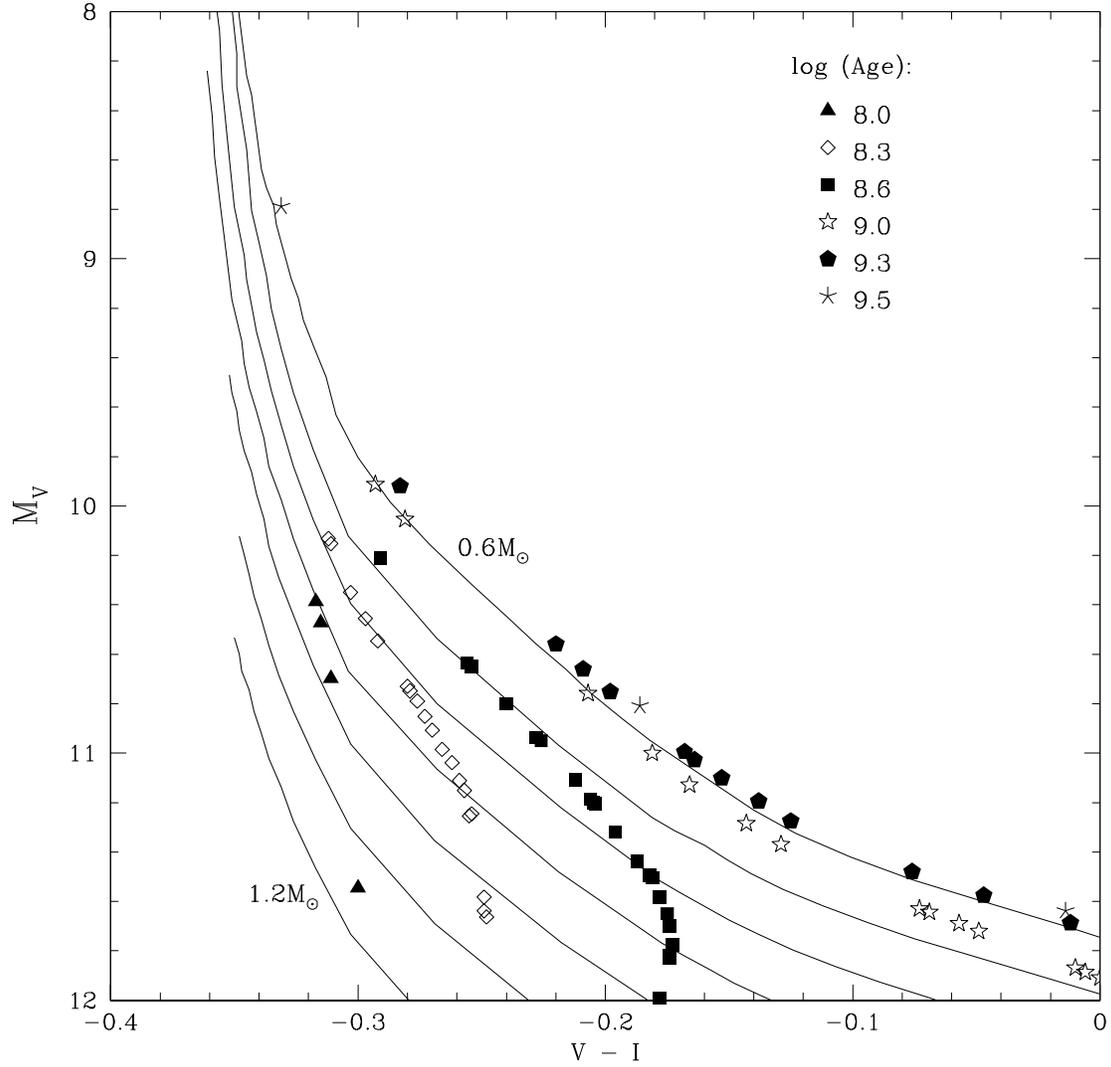}
	\caption{WD region of the CMDs from Figure 1, with cooling tracks for 
		 constant WD mass overlaid.  As age increases from $10^{8}$ 
		 years (solid triangles) to $3\times10^{9}$ years (asterisks), 
		 WDs follow tracks of lower and lower mass.}
	\label{zoom_wd}
\end{figure}

   In Figure \ref{zoom_wd} we plot just the expanded WD region of Figure 
\ref{simclus}.  Overlaid in this figure are cooling tracks of constant WD 
mass, i.e., the cooling track of an individual WD with a given (fixed) mass, 
plotted here for several masses.  As illustrated in the figure, the WD 
isochrones (represented here by various symbols) make a cut across a different 
combination of WD cooling tracks depending on the cluster's age, giving rise 
to differences in slope and position relative to the MS.  As age increases, 
WDs follow successively lower mass cooling tracks.  These differences are 
what we exploit to derive ages from just the bright cluster WDs.

   To state it another way, given a fixed magnitude range, the range of WD 
masses varies depending on the cluster's age.  In younger clusters, the mass 
range is much wider in the bright WD regime than it is in older clusters.  
This mass dependence is the cause of the observed curvature in the WD cooling 
sequence.  The effect is a direct consequence of the IFMR, the mass-radius 
relation for WDs, and the effect of the Stefan-Boltzmann law in the CMD.

\subsection{Testing the Bright WD Idea}
  \label{test_details}

   In order to explore the age information available in the photometry of the 
bright cluster WDs, we simulated open cluster CMDs, removed the stars that 
have been traditionally used for their age information, i.e., the turn off 
stars and the faintest WDs, and then analyzed the remaining, incomplete CMDs 
with a new Bayesian algorithm, currently under development by our group 
\citep{ted06}.  This process is outlined in detail below.

   First, we simulated a cluster with a given age, metallicity, distance, and
reddening.  The simulations incorporate a \citet{miller79} initial mass
function (IMF), MS and giant branch stellar evolution timescales of 
\citet{girardi00}, the IFMR of \citet{weidemann00}, WD cooling timescales of 
\citet{wood92}, and WD atmospheres colors from \citet{berg95}.  It should be 
noted that other model ingredients could be used, e.g., a different IMF or a 
different WD cooling model, but our results would not change since the 
morphology of the bright WD region in the CMD would remain essentially the 
same.  Additionally we note that all simulated cluster WDs used here are 
hydrogen-rich, DA WDs.  This is an adequate assumption; while 7\% of the 
field WDs are DBs \citep{kleinman04}, no DBs have been found in open clusters 
to date \citep{kalirai05}.  Although it should be noted that 
\citet{williams06} have recently presented the discovery of a hot DQ (a WD with a 
He-dominated atmosphere with opacity dominated by atomic carbon) in the open 
cluster M35.  However, even with this discovery, DAs still overwhelmingly 
dominate non-DAs in open clusters, making our assumption to exclude non-DAs in 
our simulations a valid approximation.

   Our models do not include effects due to residual nuclear burning in the 
surface layers of the WD.  \citet{ibenmac86} explored the effects of the mass 
of the hydrogen surface layer on the luminosity of the WD.  They found that 
the luminosity from nuclear reactions ($L_{CN} + L_{pp}$) adds very little to 
the overall luminosity of the star.  As the WD cools, the effect is less than 
or of order 10\%.  Slight systematic changes to the shape of the WD cooling 
sequence due to this effect will be higher order for the relative ages we are 
deriving.  As we calibrate the bright WD method to obtain absolute ages, our 
observations of hot cluster WDs will help us further understand exactly how 
residual nuclear burning affects the shape of the WD cooling sequence.

%will cause systematic errors that will not affect the 
%relative ages we are deriving.  Thoroughly understanding the consequences of 
%residual hydrogen burningit will be necessary to better 
%understand this effect and its consequences in affecting the shape of the WD 
%cooling sequence in the CMD.  

   Additionally, we have not included field stars in our simulations thus 
far.  This is done mainly for simplicity in the early exploration of the new 
technique.  As we continue to explore the technique and further our 
simulations, we will incorporate field stars.
                                                                                
%   The shape of the WD cooling sequence in the regime of the hot WDs may be 
%affected by residual hydrogen burning.  \citet{ibenmac86} explored the 
%effects of the mass of the hydrogen surface layer on the luminosity of the WD 
%and found that residual hydrogen burning adds very little to the overall 
%luminosity of the star.  At most, the effect is $\approx$ 10\%, but for most of 
%the cooling time of the WD, it is much less.

%Especially in the early stages of the cooling of the WD, the 
%luminosity due to hydrogen burning is many magnitudes less than .

%   Uncertainties in the physics of the hot WDs may cause one to wonder about 
%various effects, besides cluster age, changing the shape of the cooling 
%sequence in this regime.  For example, what effect does residual hydrogen 
%buring have on the shape of the cooling sequence?  Iben and MacDonald (1986) 
%have explored this question.  Their results demonstrate that residual hydrogen 
%burning adds very little to the overall luminostity of the WD.

%   Because we are only considering relative ages, it is important to note that 
%various effects that introduce possible systematic errors are 

   After we simulated a cluster, we introduced observational scatter and CMD 
incompleteness into the simulated cluster.  For a given limiting S/N, we 
calculated Gaussian photometric errors for each star in the CMD.  Various 
values for the S/N and lower $M_{V}$ (cutoff) were used to allow us to 
determine the sensitivity of the technique to photometric errors and the level 
of completeness required to still derive meaningful ages.  We will discuss 
these results more extensively in Section \ref{results}.  We also imposed an 
upper $M_{V}$ (cutoff) = 6 to ensure that our Bayesian algorithm was not able 
to derive any age information from the MSTO.

   Finally, we applied our Bayesian technique to the simulated, scattered, and
incomplete CMD.  Based on this, MCMC sampled the posterior distribution of the
age of the cluster (in log space), as well as the posterior distributions of 
other cluster parameters, namely metallicity, distance, and reddening.  Here, 
MCMC was set to run for 1,000,000 iterations.  The burn in period, that is, 
the time it takes MCMC to stabilize, was typically 20,000 iterations. Once the 
posterior distribution of the cluster's age was sufficiently sampled, we 
calculated the mean and standard deviation of the posterior distribution and 
compared these statistics  with the known age from the original simulation.  
(All statistics were calculated using values after the burn in period.)

   To understand the dependence of this new technique on various factors 
(e.g., S/N requirements, CMD completeness requirements, number of WDs 
required), we ran tests for clusters of several different ages with different 
photometric precision, CMD completeness, varying number of cluster members, 
etc.  The simulated cluster ages ranging from log (age) = 8.3 to 9.5 (i.e., 
0.2 Gyr to 3 Gyr); all clusters assumed no reddening (i.e., $A_{V}$ = 0), a 
distance modulus of 0.0\footnotemark\footnotetext{The distance modulus we 
chose to use is arbitrary since we apply photometric errors to stars as a 
function of their absolute magnitude.  We chose (m-M) = 0.0 so we can 
consistently refer to absolute, rather than apparent, magnitude.}, and solar 
metallicity.  In the following section we describe each point tested in detail.

\section{Results}
  \label{results}

   Because we set the age of each simulated cluster, it is straightforward to 
compare the precision of the MCMC output as a function of various parameters, 
including limiting S/N, number of WDs, and input precision of other cluster 
parameters.  We discuss each of these points individually below.

   Note that throughout we are discussing the precision of the ages obtained 
with the bright WDs rather than the accuracy.  We utilize the morphology of 
the bright WDs relative to the MS, and as a result, we are not claiming 
absolute, externally accurate ages.  Before that is possible, a calibration of 
this technique with other techniques (e.g., MSTO ages) must be performed.

\subsection{S/N Requirement}
  \label{sn}

\begin{figure}
   \epsscale{1}
	\plotone{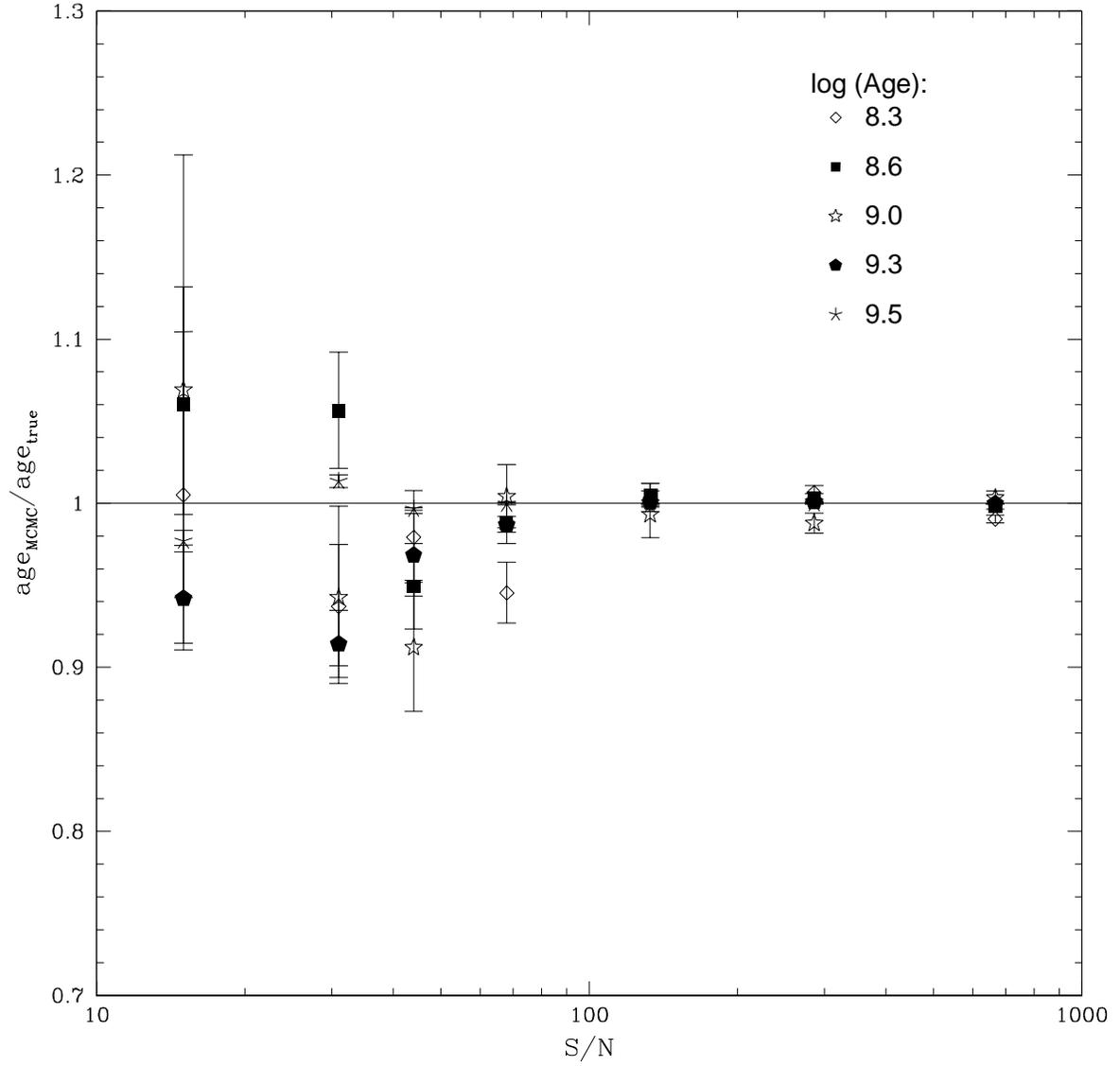}
        \caption{Relationship between the true cluster age ($age_{true}$), 
		 the mean of the age distribution obtained via MCMC 
		 ($age_{true}$), and the S/N for $M_{V}$ (cutoff) = 12.  The 
		 error bars represent the standard deviation in the posterior 
		 distribution.}
        \label{snplot}
\end{figure}

   We performed tests of the bright WD idea with several limiting S/N levels.  
This was done in order to understand the dependence of age precision on 
limiting S/N, and to test if it is possible to obtain acceptable age precision 
with realistically achievable S/N levels.  For clusters of log (age) = 8.3, 8.6, 
9.0, 9.3, and 9.5, and given input values of [Fe/H] = 0.0 $\pm$ 0.3 dex, $A_{V}$ 
= 0.0 to 0.2 magnitudes, ($m-M$) = 0.0 $\pm$ 0.1 magnitudes, and an $M_{V}$ 
(cutoff) = 12, we test age precision versus S/N in Figure \ref{snplot}.  The 
number of WDs in each cluster varied due to the stochastic nature of the 
simulations.

   As expected, precision in age results improved as the S/N increased.  These 
results are encouraging, as it demonstrates that age precision of 20\% can be 
achieved at even the lowest S/N we tested, here S/N = 15, and age precision of 
10\% is achievable by only modestly improving the S/N to 30.  Because 
achieving photometric S/N levels $\geq$ 100 are difficult, we limited further 
testing to S/N levels from $\approx$ 30 to 70\footnotemark\footnotetext{
Because achieving photometric accuracy of better than 1--2\% is very 
difficult, we discuss {\it internal} precision only.  The bright WD technique 
is constrained by the position of the WD sequence relative to the MS; and 
because absolute photometric errors cause all stars within the CMD to suffer 
the same offset, the relative position of the WD sequence and the MS will 
remain the same.  Thus, the technique relies most heavily on internal 
precision rather than external accuracy.  Nonetheless, we do not focus on 
precision better than achievable accuracy levels.}.

\subsection{Required Number of WDs}
   \label{nwds}

\begin{figure}
   \epsscale{1}
        \plotone{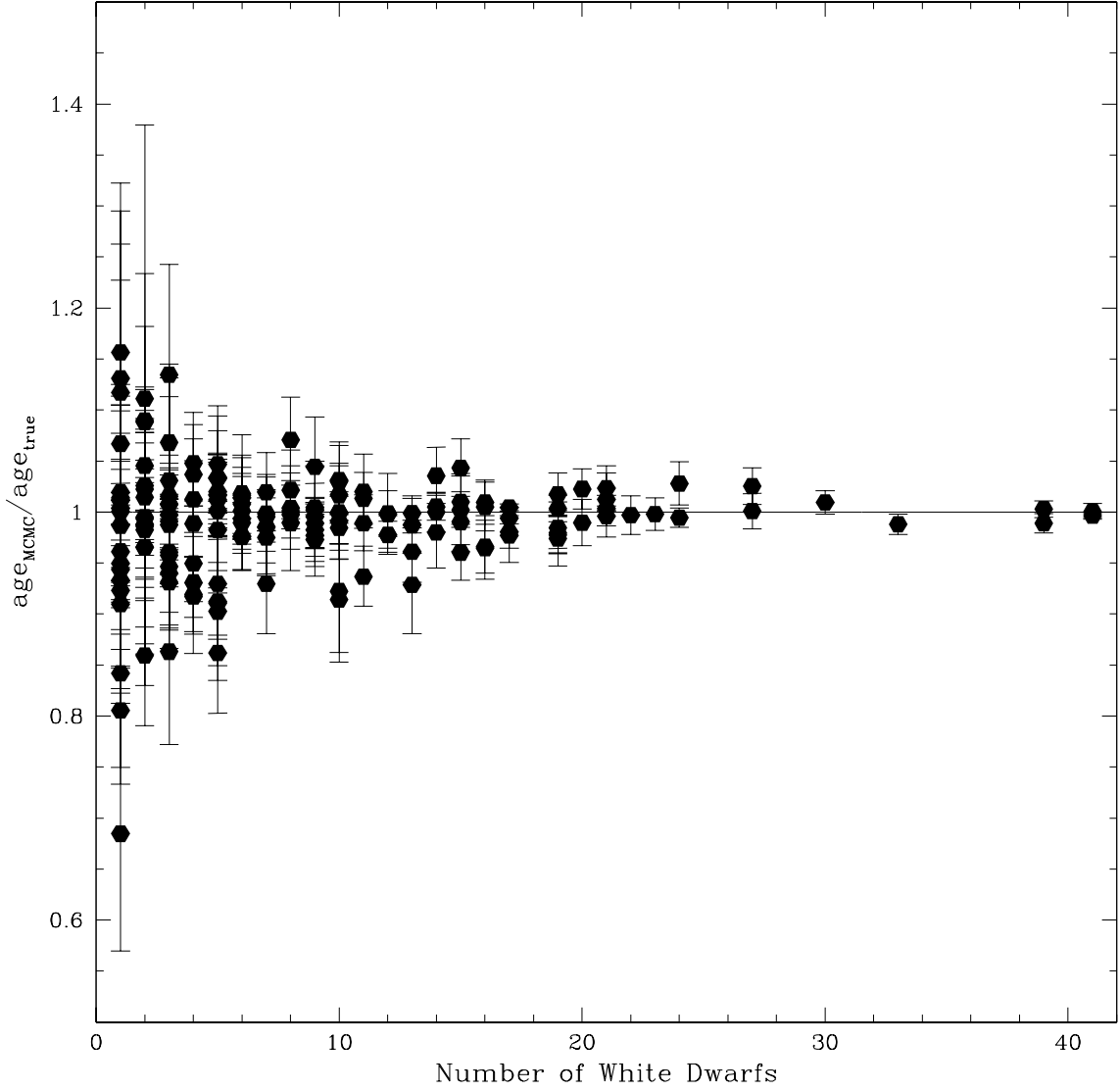}
        \caption{Relationship between the true cluster age, the mean of the
                 age distribution obtained via MCMC, and the number of WDs
                 observed.}
        \label{nwdsplot}
\end{figure}

   How many WDs are required for the bright WD technique to yield useful 
results?  To test the dependence of our new technique on the number of bright 
WDs, we simulated several clusters (all with an age of 1 Gyr and S/N = 45 at 
$M_{V}$ (cutoff) = 12) with a varying number of WDs populating the hot end of 
the WD cooling sequence.  We present results of the precision in age versus 
the number of WDs in Figure \ref{nwdsplot}.

   As evident in Figure \ref{nwdsplot}, the greater the number of WDs, the 
better the age precision.  These results show that an age precision of 10\% 
can be determined with as few as four WDs.  

\subsection{CMD Completeness Requirement}
  \label{Vcutoff}

   So far we have tested the bright WD technique assuming $M_{V}$ (cutoff) = 
12.  However, when obtaining real data, the WD cooling sequence may be 
either more or less complete.  We therefore explore the dependence of age 
precision on CMD completeness.  To test this, we simulated several clusters of 
various ages, imposing incompleteness at $M_{V}$ = 11, 12, and 13 with a 
photometric S/N of 45 at $M_{V}$ (cutoff).  Again, the number of WDs in each 
cluster varied based on the individual simulations.

   We show the results in the top panel (A) of Figure \ref{other_plots}.  The  
plotted points represent the averages of the results of many individual 
simulations in that particular location of parameter space; the error bars 
are the standard deviations of the results.   As expected, as the $M_{V}$ 
(cutoff) becomes fainter and more WDs are included, the age precision 
increases.  For clusters of 1 Gyr, $M_{V}$ (cutoff) = 12 (13) typically yields 
an age precision of $\leq$ 10\% (5\%).  For younger clusters good results can 
be achieved even at $M_{V}$ (cutoff) = 11, e.g., for log (age) = 8.6, the 
typical age precision is 10\%.

\subsection{Precision of [Fe/H] and Other Priors}
  \label{params}

   The Bayesian technique requires prior information on inputs such as 
distance, metallicity, and reddening.  How sensitive are the results to these 
priors?  That is, how precisely do we need to know cluster metallicity, 
distance, or reddening for the bright WD technique to be useful?  In addition 
to age, the method recovers posterior distributions of metallicity, distance, 
and reddening.  The widths of the posterior distributions of these quantities 
are substantially narrower than the priors, indicating that all the requisite 
information is contained in the CMD.

   For example, we simulated clusters with an age of 1 Gyr and an $M_{V}$ 
(cutoff) = 12.  We chose metallicity precisions of 0.1, 0.2, 0.3, 0.4, and 0.5 
dex.  We display the relationship between age precision and precision of input 
metallicity in the middle panel (B) of Figure \ref{other_plots}.  As in the 
top panel of the figure, the points represent the averages of age results from 
many simulations with the standard deviations represented by the error bars.  
Results show little, if any, dependence between the prior precision on 
metallicity and the age determined from the bright WDs.  This does not mean 
that cluster metallicities are unimportant in this work.  On the contrary, we 
hope to eventually test cluster MS ages versus cluster WD ages at a range of 
metallicities.  Rather, good cluster metallicities are not required for a 
precise bright WD age.

\subsection{Age Range}
   \label{agereq}

\begin{figure}
   \epsscale{1.00}
        \plotone{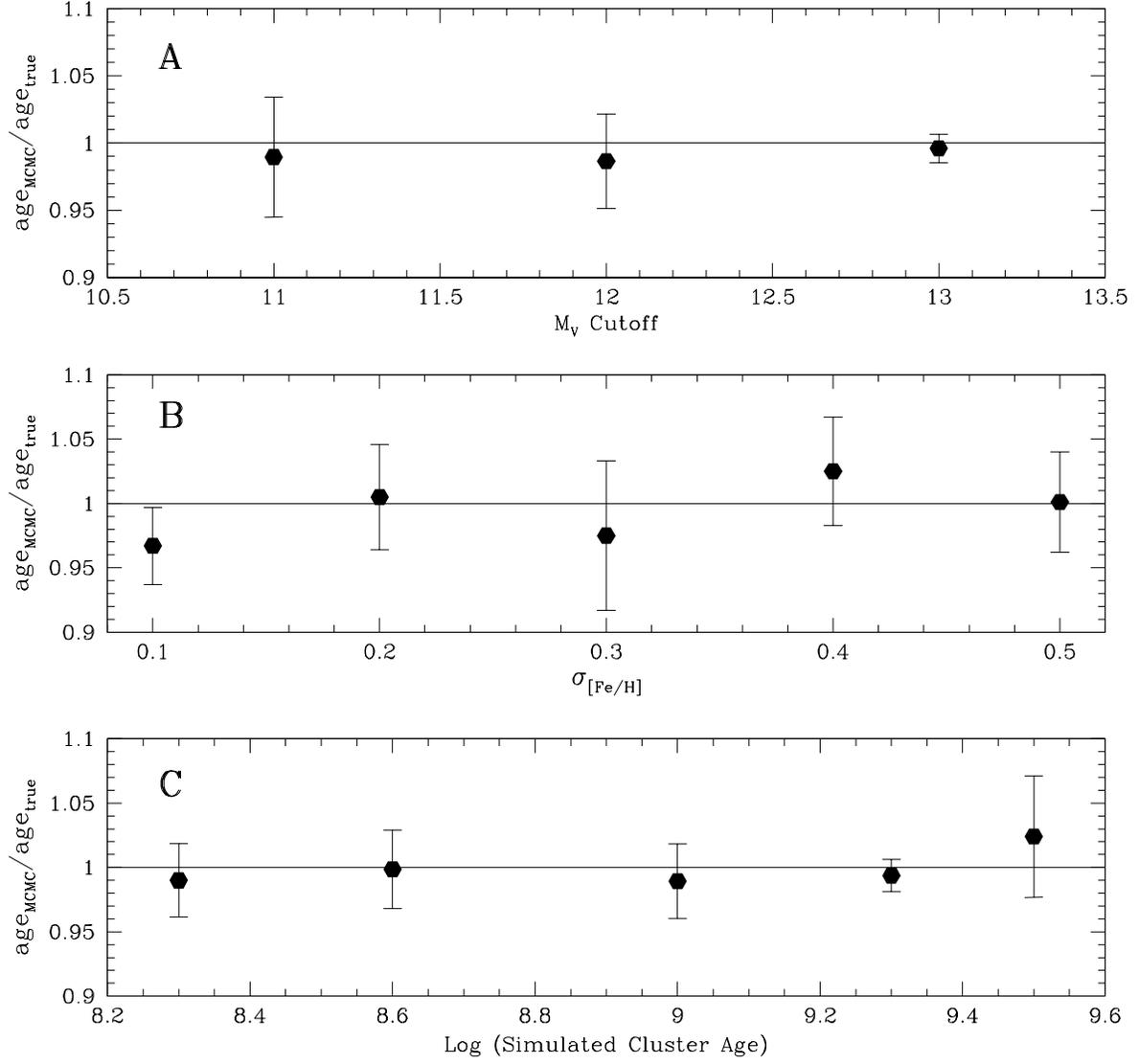}
        \caption{Relationship between the true cluster age, the mean of the
                 age distribution obtained via MCMC, and (A) the CMD 
		 completeness, (B) prior precision of the cluster metallicity, 
		 and (C) the simulated cluster age.}
	\label{other_plots}
\end{figure}

   As mentioned in Section \ref{rationale}, as a cluster ages, the slope of its 
WD cooling sequence becomes increasingly indistinguishable from other old 
clusters.  Because of this, we desired to understand how the precision of 
the bright WD age varied as a function of increasing age.  We show the  
results of this in panel C of Figure \ref{other_plots}.  We have not yet 
tested any ages older than log (age) = 9.5 (i.e., 3 Gyr) due to limitations in 
our input models; however, as we incorporate more high mass WD cooling models, 
we will continue to push this technique to greater ages.

   We display the results in the bottom panel (C) of Figure 
\ref{other_plots}.  As usual, the points are the averages of several 
simulations with the standard deviations given by the error bars.  Based on 
this figure, we see that for the ages tested, the age precision of the bright 
WD technique does not significantly decrease with increasing age.  Further 
testing will be done to determine this effect for clusters of ages above 3 
Gyr, placing an important constraint on the technique.

   As the age of a cluster increases, two main factors contribute to the age 
precision.  The first is that as the cluster becomes older, there are more 
WDs.  As demonstrated in Section \ref{nwds} (particularly Figure 
\ref{nwdsplot}), as the number of WDs increases, so does the age precision.  
Competing with this effect is that as the cluster ages, the bright portion of 
its WD cooling sequence becomes increasingly difficult to distinguish from 
other old clusters (see Figures \ref{simclus} and \ref{zoom_wd}).  It is 
possible that there exists a ``sweet spot," where these two effects complement 
each other and contribute to a very precise age.  This may be the cause of the 
sudden tightness in age precision around log (age) = 9.3 in Figure 
\ref{other_plots}.  However, as a cluster increases in age, the number of 
WDs quickly begins to do little to improve age precision while the WD 
sequences become more and more degenerate.  Further testing, particularly 
expanding tests out to older clusters, will more clearly help us understand 
this effect and constrain the bright WD technique.

\section{Conclusions}
  \label{conclusion}

   Current observational techniques to obtain ages from cluster WDs are 
challenging, due to the faintness of the coolest cluster WDs.  We have shown 
that the bright WDs can be used to determine cluster ages.  This is done by 
exploiting the differences in slope and position relative to the MS of the WD 
cooling sequence in the regime of the bright WDs caused by varying WD mass 
distributions.  By employing a new Bayesian technique to fully extract all age 
information in the WD portion of the CMD, we have shown that there is 
sufficient age information in the bright WDs.  

   With the assumption of a single-valued IFMR, our studies show that we can 
achieve age precision of 10\% with a S/N $\geq$ 30 at $M_{V}$ (cutoff) = 12, 
with as few as 4 WDs for low reddening clusters.  We find no dependence of age 
precision on the prior precision of metallicity, distance, and reddening, nor 
on the age of the cluster, at least up to the oldest age (3 Gyr) tested in 
this study.  Additional studies will be done to determine if this technique is 
feasible for even older ages, particularly ages comparable to those of the 
globular clusters.

   If the bright WDs technique continues to be successful in determining 
precise cluster ages, WD cosmochronometry can be applied to more distant 
and/or older clusters (where observing to the WD terminus is especially 
challenging).  This will increase the sample of objects available for 
study, particularly allowing us to sample clusters in age-metallicity 
space that would be otherwise too difficult using the current method of 
determining WD ages.  This technique will allow WDs to be an increasingly 
powerful and effective tool for studying the ages of stellar populations.

\begin{acknowledgements}

   This material is based upon work supported by the National Aeronautics and 
Space Administration under Grant No.\ NAG5-13070 issued through the Office of 
Space Science, and by the National Science Foundation through Grant 
AST-0307315.

\end{acknowledgements}

%% Figure 1

%% Figure 2

%% Figure 3

%% Figure 4

%% Figure 5


\begin{thebibliography}{}

\bibitem[Bergeron et al.(1995)]{berg95}Bergeron, P., Wesemael, F., \& Beauchamp, A. 1995, \pasp, 107, 1047
\bibitem[Chaboyer et al.(1996)]{chab96}Chaboyer, B., Demarque, P., \& Sarajedini, A. 1996, \apj, 459, 558
\bibitem[Claver(1995)]{clav95}Claver, C. F. 1995, PhD Thesis, The University of Texas at Austin
\bibitem[Claver et al.(2002)]{clav02}Claver, C. F., Liebert, J., Bergeron, P. 2001, \apj, 563, 987
\bibitem[Girardi et al.(2000)]{girardi00}Girardi, L., Bressan, A., Bertelli, G., \& Chiosi, C. 2000. A\&AS, 141, 371
\bibitem[Iben and MacDonald(1986)]{ibenmac86}Iben, I. \& MacDonald, J. 1986, \apj, 301,164
\bibitem[Kleinman et al.(2004)]{kleinman04}Kleinman, S.J., et al. 2004, \apj, 607, 426
\bibitem[Kalirai et al.(2005)]{kalirai05}Kalirai, J.S., Richer, H.B., Hansen, B.M.S., Reitzel, D., \& Rich, R.M. 2005, \apj, 618, L129
\bibitem[Knox et al.(1999)]{knox99}Knox, R.A., Hawkins, M.R.S., \& Hambly, N.C.
1999, \mnras, 306, 736
\bibitem[Leggett et al.(1998)]{leggett98}Leggett, S.K., Ruiz, M.T., \& Bergeron, P. 1998, \apj, 497, 294
\bibitem[Miller \& Scalo(1979)]{miller79}Miller, G. E., \& Scalo, J. M. 1979, \apjs, 41, 513
\bibitem[Oswalt et al.(1996)]{oswalt96}Oswalt, T.D., Smith, J.A., Wood, M.A., \& Hintzen, P. 1996, \nat, 382, 692
\bibitem[Richer et al.(1998)]{richer98}Richer, H.B., Fahlman, G.G., Rosvick, J., \& Ibata, R. 1998, \apj, 504, L591
\bibitem[von Hippel(2001)]{vh01}von Hippel, T. 2001, in ASP Conf. Ser. 245, ed. T. von Hippel et al., (San Fransisco: ASP), 190
\bibitem[von Hippel(2005)]{ted05}von Hippel, T. 2005, \apj, 622, 565
\bibitem[von Hippel \& Gilmore(2000)]{vg00}von Hippel, T., Gilmore, G. 2000, \aj, 120, 1384
\bibitem[von Hippel, Gilmore, \& Jones(1995)]{vgj95}von Hippel, T., Gilmore, G., \& Jones, D.P.H. 1995, \mnras, 273, L39
\bibitem[von Hippel et al.(2006)]{ted06}von Hippel, T., Jefferys, W. H., Scott,
J., Stein, N., Winget, D. E., DeGennaro, S., Dam, A., \& Jeffery, E. J. 2006, \apj, 645, 1436
\bibitem[Weidemann(2000)]{weidemann00}Weidemann, V. 2000, A\&A, 363, 647
\bibitem[Williams et al.(2006)]{williams06}Williams, K. A., Liebert, J., Bolte, M., Hanson, R. B. \apj, 643, L127
\bibitem[Winget et al.(1987)]{winget87}Winget, D. E., Hansen, C. J., Liebert, J., van Horn, H. M., Fontaine, G., Nather, R. E., Kepler, S. O., \& Lamb, D. Q. 1987, \apj, 315, L77
\bibitem[Wood(1992)]{wood92}Wood, M. A. 1992, \apj, 386, 539

\end{thebibliography}
\end{document}